\documentclass[aps, prl, 10pt, twocolumn,superscriptaddress, longbibliography, nobalancelastpage]{revtex4-1}

\usepackage{graphicx}
\usepackage{graphics}
\usepackage{amsmath}
\usepackage{amssymb}
\usepackage{amsfonts}
\usepackage{dsfont}
\usepackage{braket}
\usepackage{color}
\usepackage{braket,slashed}
\usepackage[mathscr]{euscript}
\definecolor{darkblue}{rgb}{0, 0, 0.8}
\usepackage[colorlinks=true, breaklinks=true, linkcolor=red, citecolor=blue, urlcolor=blue]{hyperref} 
\usepackage{subfigure}
\usepackage{xfrac}
\usepackage{bm}
\usepackage{kantlipsum}
\usepackage{enumitem}
\usepackage{tikz}
\usepackage{framed}
\usepackage{graphicx}
\usepackage{subfigure}
\usepackage{lipsum}

\allowdisplaybreaks[1]

\newcommand{\code}[1]{\texttt{#1}}

\begin{document}

\title{A scaling hypothesis for matrix product states}

\author{Bram Vanhecke}
\affiliation{Department of Physics and Astronomy, University of Ghent, Krijgslaan 281, 9000 Gent, Belgium}

\author{Jutho Haegeman}
\affiliation{Department of Physics and Astronomy, University of Ghent, Krijgslaan 281, 9000 Gent, Belgium}

\author{Karel Van Acoleyen}
\affiliation{Department of Physics and Astronomy, University of Ghent, Krijgslaan 281, 9000 Gent, Belgium}

\author{Laurens Vanderstraeten}
\affiliation{Department of Physics and Astronomy, University of Ghent, Krijgslaan 281, 9000 Gent, Belgium}

\author{Frank Verstraete}
\affiliation{Department of Physics and Astronomy, University of Ghent, Krijgslaan 281, 9000 Gent, Belgium}

\begin{abstract}
We revisit the question of describing critical spin systems and field theories using matrix product states, and formulate a scaling hypothesis in terms of operators, eigenvalues of the transfer matrix, and lattice spacing in the case of field theories. Critical exponents and central charge are determined by optimizing the exponents such as to obtain a data collapse. We benchmark this method by studying critical Ising and Potts models, where we also obtain a scaling ansatz for the correlation length and entanglement entropy. The formulation of those scaling functions turns out to be crucial for studying critical quantum field theories on the lattice. For the case of $\lambda\phi^4$ with mass $\mu^2$ and lattice spacing $a$, we demonstrate a double data collapse for the correlation length $ \delta \xi(\mu,\lambda,D)=\tilde{\xi} \left((\alpha-\alpha_c)(\delta/a)^{-1/\nu}\right)$ with $D$ the bond dimension, $\delta$ the gap between eigenvalues of the transfer matrix, and $\alpha_c=\mu_R^2/\lambda$ the parameter which fixes the critical quantum field theory. 
\end{abstract}

\maketitle

\noindent\emph{Introduction.} Traditional numerical techniques for simulating extensive many-body systems such as Monte-Carlo sampling and exact diagonalizations naturally come with a dimensionfull parameter that controls the level of approximation: the system size $L$. One of the major insights that has allowed the simulation of systems at or near criticality has been the realisation that, rather than simply pushing $L\rightarrow \infty$, the numerical results at different system sizes can be combined in a much more informed way using the concept of finite-size scaling \cite{Fisher1972, Brezin1982, Cardy1988}. The crucial idea is that $1/L$ acts as a relevant perturbation away from criticality, and thus enters in the singular part of the free energy. By invoking the scaling hypothesis, the latter is a generalized homogeneous function
\begin{equation}
f(\{t_i\},1/L)= s^{-d} f( \{s^{\alpha_i} t_i\} ,s/L) \;,
\end{equation}
where $\{t_i\}$ is the set of coupling constant corresponding to (relevant) perturbations away from criticality in the theory. This scaling hypothesis can then be used to perform a collapse of numerical data sets for different $L$, and as such to obtain accurate estimates for critical exponents and the location of a critical point. Furthermore, in the context of quantum field theory (QFT) lattice simulations, finite-size scaling ideas have proven vital for reaching the continuum limit \cite{Luscher:1991wu,Jansen:1995ck}.    

Here we consider the application of matrix product states (MPS) methods \cite{White1992, Verstraete2004, Verstraete2008, Schollwoeck2011} for simulating critical 1D quantum or 2D classical spin systems, including the continuum limit of lattice descriptions for QFTs. In particular we will study uniform MPS that, in contrast to the techniques that we discussed above, work directly in the thermodynamic limit $L\rightarrow \infty$. In this case the level of approximation is set by the finite bond dimension $D$ of the matrices, which is in fact a proxy for the finite amount of entanglement in the simulated state. The area law for the entanglement entropy \cite{Hastings2007} validates this approximation and explains the success of MPS methods for parameterizing ground states of gapped systems \cite{Hastings2007, Verstraete2006}. Relying on this success, the traditional approach to study phase transitions, and, relatedly, continuum limits of lattice field theories, with MPS has been to extrapolate MPS predictions for one or more order parameters $M'=\lim_{D\rightarrow\infty}M(D)$ towards the limit of infinite bond dimension $D$. By then studying the behaviour of $M'$ throughout the phase diagram, the location of the critical point and critical exponents can be estimated. Analogous to the effect of a finite size, it has been recognized that the finite bond dimension of an MPS acts as a relevant perturbation and induces an additional length scale in the problem, that shows crossover behaviour with the finite system size \cite{nishino1996numerical, Pirvu2012}. MPS methods can however work directly in the thermodynamic limit, such that, at criticality, only the length scale associated with the finite bond dimensions remains. This length scale was identified as the correlation length and was shown to scale as $\xi^{(D)} \sim D^\kappa$ in the asymptotic limit of large $D$ \cite{Tagliacozzo2008}. Here, a new critical exponent $\kappa$ was introduced, the value of which is completely specified by the central charge of the underlying conformal field theory \cite{Pollmann2009, Pirvu2012}. 

In light  of the success of the aforementioned finite-size scaling methods, proper finite-entanglement scaling ans\"atze are paramount for the further development of the tensor network framework for simulating (near) critical theories and QFTs. This holds even more for higher dimensional systems simulated with the generalization of MPS, known as projected entangled pair states (PEPS) \cite{Verstraete2004b}, for which the computational cost grows much faster with increasing bond dimension $D$. The scaling behavior in the discrete variable $D$ only holds for sufficiently large $D$ and cannot be expected to be smooth or homogeneous for small or intermediate values of $D$. Instead, scaling ans\"atze can be formulated directly in terms of the finite correlation length $\xi^{(D)}$ \cite{Corboz2018, Rader2018, Czarnik2019, Pillay2019}. However, the (inverse) correlation length $1/\xi^{(D)}$ only correctly quantifies the strength of the relevant perturbation at criticality, when no other relevant perturbations are present, as otherwise it does not tend to zero for $D\to\infty$.

In this paper, we motivate and introduce the use for entanglement scaling of a different (inverse) length scale $\delta$ defined in terms of the gaps in the full spectrum of (inverse) correlation lengths, as obtained from the (negative) logarithm of the eigenvalues of the transfer matrix \cite{Zauner2015}. A careful study of the nature of the MPS approximation \cite{Zauner2015, Rams2015, Bal2016} indicates that these gaps are a direct consequence of the finite bond dimension and go to zero for $D\to \infty$, regardless whether the system is gapped or not. In particular, the correlation length $\xi^{(D)}$ can itself be scaled in terms of $\delta$, as was first illustrated by Rams et al for gapped systems \cite{Rams2018}. Combined with other relevant perturbations, a full scaling ansatz for the correlation length itself can thus be formulated. We illustrate this for the two-dimensional classical Ising and Potts models. Understanding the resulting scaling functions is also of crucial importance to simulate continuum limits of spin systems in the form of QFT's. The continuum limit is obtained by taking the limit of bond dimension going to infinity and lattice spacing to zero, and we demonstrate that this double scaling limit yields a double data collapse near the critical point of (2+0)-dimensional  $\lambda\phi^4$ theory.\vspace{0.2cm}

\noindent \emph{Entanglement scaling hypothesis.}
Throughout this paper, we use uniform MPS which depend on a single tensor to parameterize translation-invariant states directly in the thermodynamic limit. An MPS with a finite bond dimension provides a variatonal approximation for low-energy states by truncating in the entanglement spectrum, which has its repercussion on the approximation of the physical properties of the system. In particular, correlation functions are represented by a linear combination of exponentially decaying functions, where the spectrum of inverse correlation lengths is determined by the eigenvalues of the MPS transfer matrix $\lambda_i$ as 
\begin{equation}
\epsilon_i = \xi_i^{-1} = - \log |\lambda_i| ,  \qquad \mathrm{with} \quad\epsilon_i \leq \epsilon_{i+1},
\end{equation}
assuming the MPS is normalized such that $\epsilon_0 = 0$. The actual correlation length of the state is then identified as $\xi \equiv \xi_1$. Close to a second order critical point, however, we expect correlation functions to exhibit a power-law contribution multiplied with the exponential decay, which can be understood from the K\"all\`en-Lehmann representation of correlation functions as a linear combination of a continuum of exponentials \cite{Zauner2015}. The MPS transfer matrix thus provides a discretized approximation to this continuous spectrum of correlation length, and the spacing between them is a reflection of an inverse system size just as in the case of the eigenvalue spacing in a finite spin chain . By interpreting the true state as resulting from an infinite amount of imaginary-time evolution (i.e.\ the path-integral representation), the discretization of the spectrum of correlation lengths can then be understood as resulting from the compression of the infinite imaginary-time interval that is inherent in the MPS approximation \cite{Rams2015, Bal2016}. Hence, the gaps in the transfer-matrix spectrum can be related to a finite size in imaginary time.

Therefore we can build a finite-entanglement scaling theory by quantifying the discreteness of the spectrum of inverse correlation lengths, i.e.\ the gaps in the transfer-matrix spectrum. The most simple definition is $\delta = \epsilon_2 - \epsilon_1$, which was indeed used in Ref.~\onlinecite{Rams2018} to extrapolate the correlation length itself. However, as also remarked in Ref.~\onlinecite{Rams2018}, the spectrum can consist out of different sectors (sometimes but not always appearing at different complex phases $\text{arg} \lambda_i = \mathrm{Im} \log \lambda_i$), and it can be useful to consider a generalised definition
\begin{equation}
\delta = \sum_{i=1}^n c_i \epsilon_i, \qquad \text{with}\quad\sum_i c_i =0,
\end{equation}
with a finite number $n$ sufficiently smaller than $D^2$, such that only the largest eigenvalues $\lambda_i$ are included. For any choice of the coefficients $c_i$ such that $\sum_i c_i =0$, this quantity should converge to zero for $D\to\infty$. Evidently, the $\epsilon_i$'s, and thus also $\delta$, transform as an inverse length under scale transformations. Therefore, we can formulate the scaling hypothesis for an order parameter $m$ as
\begin{equation}
m(t,\delta) = s^{-\beta/\nu} m \left( s^{1/\nu} t,s\delta\right).
\end{equation}
This yields the expression for a corresponding scaling function $\tilde{m}$
\begin{equation}
\delta^{-\beta/\nu} m(t,\delta)=m\left(\delta^{-1/\nu}t,1\right) = \tilde{m} \left( \delta^{-1/\nu}t \right).
\end{equation}
From the scaling property of $\delta$, it follows that the scaling functions $\tilde{m}$ away from the origin exhibit a power-law behaviour with the exponent corresponding to the operator one is looking at. However, unlike in traditional finite-size scaling, where the finite size imposes smoothness on the scaling function, the behaviour of the scaling function around the origin, i.e.\ in the vicinity of the critical point or for large $\delta$ (small bond dimension), reproduces mean-field behaviour, consistent with the findings of Ref.~\onlinecite{liu2010symmetry}. As a consequence, the scaling function can still be non-analytic, but exhibits the mean-field exponents around the origin.

\begin{figure*} \begin{center}
\subfigure{\includegraphics[width=0.68\columnwidth]{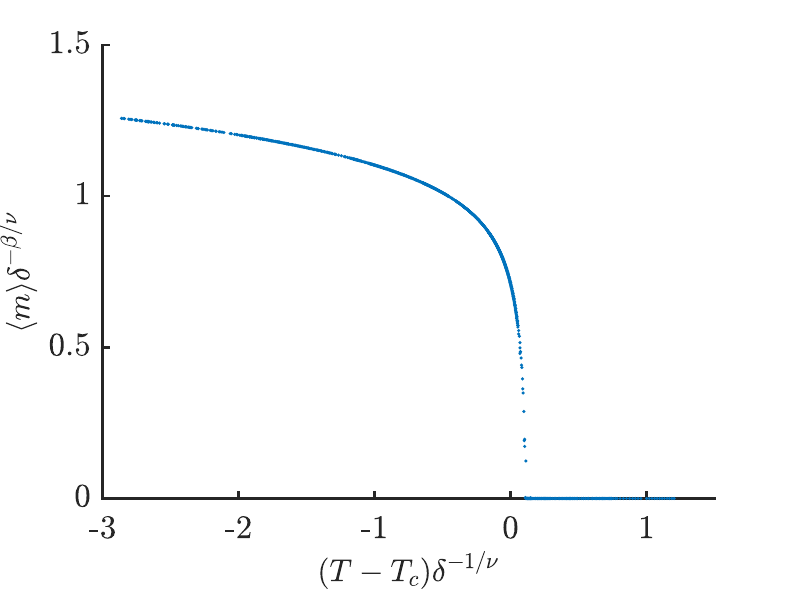}}
\subfigure{\includegraphics[width=0.68\columnwidth]{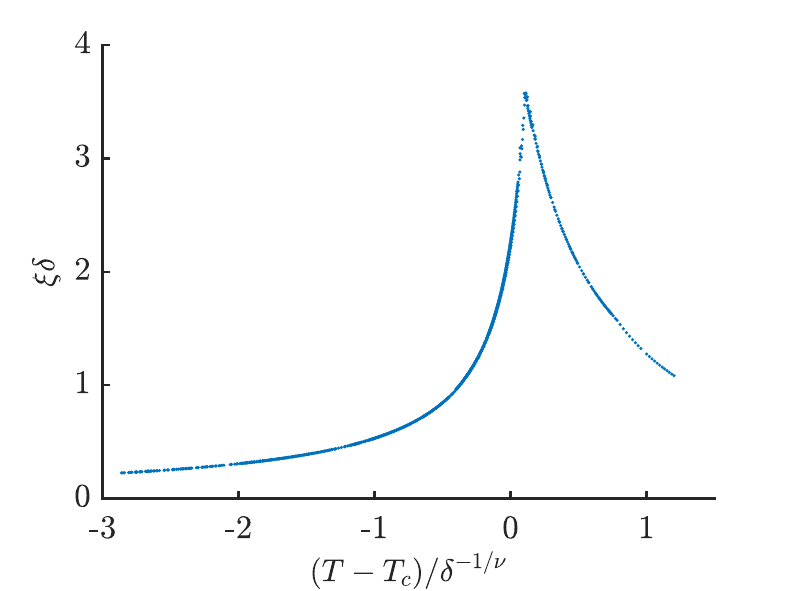}}
\subfigure{\includegraphics[width=0.68\columnwidth]{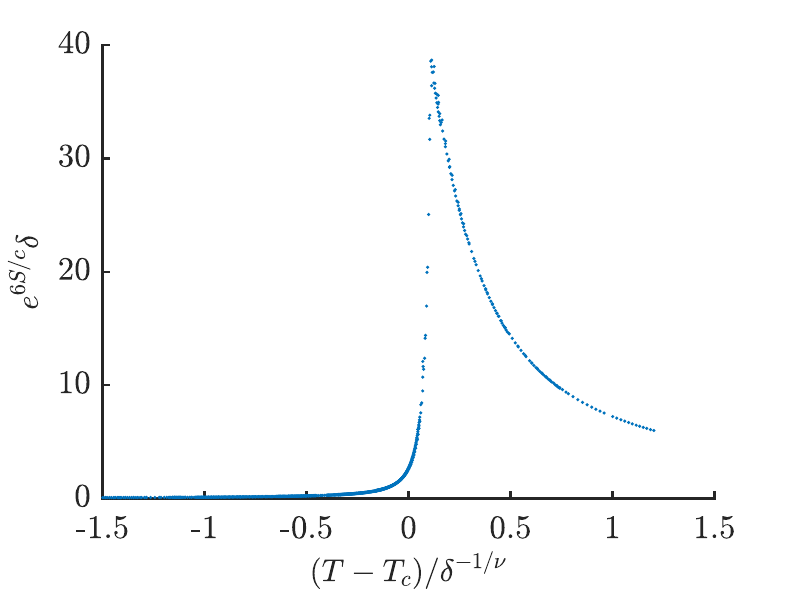}}
\caption{Collapse plots for the Ising model, calculated with MPS of bond dimension 12,20,30,50,90 and 150, for 189 different temperatures linearly spaced between $T=2.2666$ and $T=2.2698$.\\}
\label{fig:ising}
\end{center} \end{figure*}

In a similar vein, we can formulate a scaling hypothesis for the correlation length
\begin{equation}
\delta \xi(t,\delta)=\xi\left(\delta^{-1/\nu}t,1\right) = \tilde{\xi}\left(\delta^{-1/\nu}t\right),
\label{eq:xicollapse}
\end{equation}
in terms of a scaling function $\tilde{\xi}$. Crucially, this scaling behaviour justifies prior approaches where a scaling ansatz for $m$ was formulated directly in terms of $1/\xi$ instead of $\delta$. But using $\delta$, which objectively quantifies the perturbation strength due to the finite bond dimension, the order-parameter scaling function $\tilde{m}$ takes a more natural form away from the origin. Furthermore, the ability to also scale the correlation length yields additional data points in order to fit more accurately the location of the critical point and the corresponding scaling exponents.

Finally, we can also extract the bipartite entanglement entropy from a given MPS. Using the CFT formula for the entanglement entropy \cite{Calabrese2004}, we know that $\exp\left(\frac{6}{c}S\right)$ scales as a length, and that we can write down a scaling hypothesis of the form
\begin{equation}
\exp\left(\frac{6}{c} S(t,\delta)\right) = s \exp \left( \frac{6}{c} S(s^{-1/\nu}t, s\delta) \right),
\end{equation}
where $c$ is the central charge for the CFT describing the critical behaviour of the model. This directly yields a scaling function for the entanglement entropy.

For a given set of data points at different MPS bond dimensions, the critical properties of the model can now be determined by optimizing a data collapse in terms of $\delta$. In principle, every $\delta$ built up from a set of $c_i$'s should give the right scaling behaviour, but in order to improve the collapse, the $c_i$'s can also be treated as parameters that can be optimized. The cost function that we optimize is the sum of the distances of all data points to a scaling function, which is itself parametrized by a set of parameters. We use standard non-linear optimization algorithms for determining these different parameters. Note that there is no consensus on an ultimate algorithm to perform finite size scaling and data collapse, and it remains an active area of research which is outside of the scope of the current paper \cite{kawashima1993critical,bhattacharjee2001measure, houdayer2004low,wenzel2008percolation,winter2008geometric}. \vspace{0.2cm}

\noindent \emph{Two-dimensional Ising and Potts models.} As a first illustration of our method, we consider the classical Ising model on the square lattice. Using the vumps algorithm \cite{Fishman2018, Vanderstraeten2019}, we have computed a set of variational MPSs of bond dimension $D$ ranging between 10 and 200, for different temperatures around the critical point $T_c=2.269185314$. Here, we fix the $c_i$'s by hand, defining $\delta=\epsilon_4-\epsilon_2$, such that we obtain a collapse of the data. In Fig.~\ref{fig:ising} we plot the scaling functions for the order parameter, correlation length and entanglement entropy, using the known values of the critical temperature, the exponents $\nu=1$ and $\beta=1/8$ and the central charge $c=1/2$. If we jointly optimize the collapse of order parameter and correlation length without prior knowledge on the critical data, we find a critical temperature of $T_c^\mathrm{MPS}=2.269184934$ with critical exponents $\nu=0.99980$ and $\beta=0.12534$. Alternatively, if we fix the exponents to their known value $\nu=1$ and $\beta=1/8$, we find $T_c^\mathrm{MPS}=2.269184914$.

\begin{figure*} \begin{center}
\subfigure{\includegraphics[width=0.68\columnwidth]{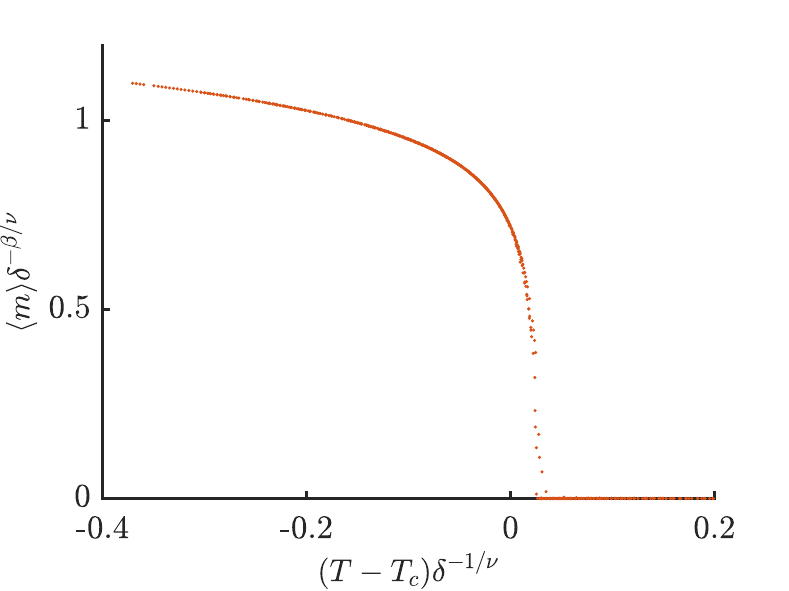}}
\subfigure{\includegraphics[width=0.68\columnwidth]{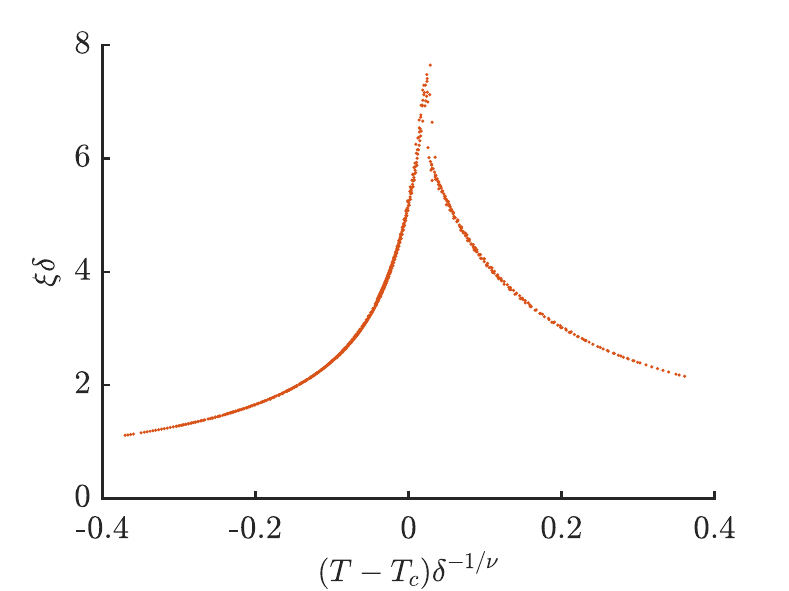}}
\subfigure{\includegraphics[width=0.68\columnwidth]{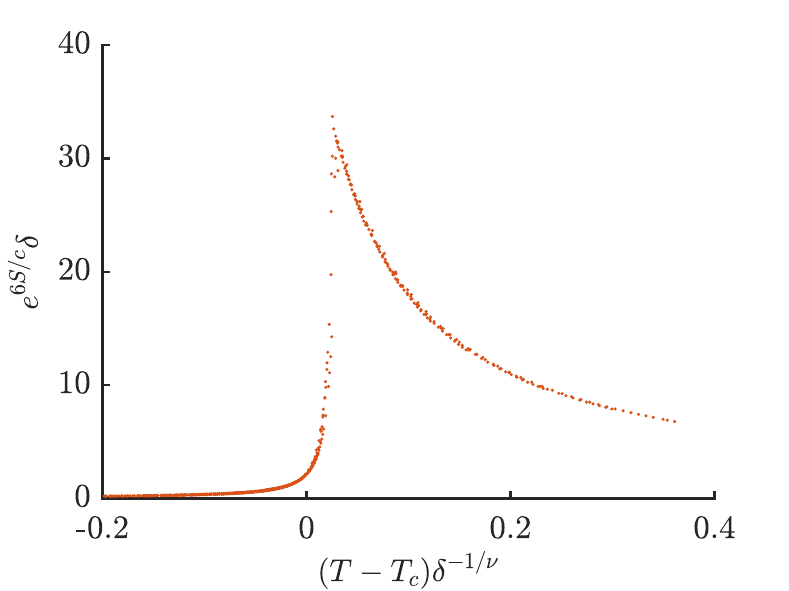}}
\caption{Collapse plots for the Potts model, calculated with MPS of bond dimension 21,31,42,50,60,81,99 and 120, for 96 different temperatures linearly spaced between $T=0.9939$ and $T=0.9954$.\\}
\label{fig:potts}
\end{center} \end{figure*}

Secondly, we study the 3-state Potts model. We have used MPS with bond dimensions $D=21\to120$ around the critical point $T_c=0.994972861$. Here, we have used $\delta=\frac{5}{2}\epsilon_2-\epsilon_4-\epsilon_5-\frac{1}{2}\epsilon_6$. Again, we plot the three scaling functions, using the known values for the critical data. If we jointly optimize the collapse of $\xi$ and $m$ for this model without prior knowledge on the critical data, we find $T_c=0.9949715$, $\nu=0.8283$ and $\beta=0.1086$, to be compared with the exact values $\nu=5/6$ and $\beta=1/9$. \vspace{0.2cm}

\noindent \emph{$\lambda \phi^4$ field theory.} Finally we look at a phase transition in a QFT as a more exotic application, described by the following lagrangian density
\begin{equation}
\mathcal{L}(\phi)=\frac{1}{2}\partial_\mu\phi\partial^\mu\phi+\frac{1}{2}\mu_p^2\phi^2+\frac{1}{4}\lambda_p\phi^4\,.
\end{equation}
The resulting Euclidean path integral can be discretised in a standard way, e.g.: $\partial_1\phi(x,y)=(\phi_{i+1,j}-\phi_{i,j})/a$ (with $a$ the lattice spacing) and converted into a tensor-network form by truncating the $\phi$-fields in a suitable basis, see Ref.~\onlinecite{Kadoh2019}. In order to study the second-order QFT phase transition from the $\mathbb{Z}_2$ unbroken phase $\braket{\phi}=0$ to the broken phase $\braket{\phi}\not=0$ \cite{Chang1976}, we computed uniform-MPS approximations of the fixed point of the path-integral transfer operator in the thermodynamic limit. In addition to the entanglement scaling $\delta\rightarrow 0$, the QFT interpretation of our numerical results then requires us to also consider the continuum scaling $a\rightarrow 0$. Rather than taking both limits separately, which up till now has been the standard procedure for MPS simulations of QFTs \cite{Banuls2013, Milsted2013, Buyens2014, Banuls2013}, we will show how one can perform a \emph{double collapse} on all the results for different lattice spacings and bond dimensions into a single scaling function. 

Let us first consider the continuum scaling of the Euclidean lattice path integral. The lattice action is defined in terms of the lattice parameters $\lambda = \lambda_p a^2$ and $\mu^2= \mu_p^2 a^2$. (We use the subscript $p$ for quantities in \emph{physical} units, independent of the lattice spacing $a$.) However, in the continuum limit, the mass term receives a divergent one-loop correction, such that the bare mass $\mu_p=\mu/a$ of the theory is parameterized in terms of a renormalized mass parameter $\mu_{p,\text{R}}=\mu_{\text{R}}/a$ as
\begin{equation}
	\mu^2 = \mu_{\text{R}}^2 - 3 \lambda A(\mu_\text{R}^2),
\end{equation}
where the one-loop contribution $A(x)$ is given in e.g.\ Ref.~\onlinecite{Kadoh2019} and diverges as $log(x)$ for small values of its argument. The $\lambda\phi^4$ theory being superrenomalizable, this is the only UV divergence and the the IR behaviour is then completely characterised by the finite ratio $\alpha  = \mu_{p,\text{R}}^2 / \lambda_p = \mu_R^2/\lambda$. In approaching the continuum limit, both the mass term and interaction term get additional UV-finite corrections. Rather than computing these in perturbation theory, we parameterise general corrections and determine the coefficients as part of the scaling analysis. Specifically, we consider the following parameterisation,
\begin{align}
\mu^2 &= \lambda g -3\lambda A(\lambda g) \nonumber \\
\alpha &= g+\lambda P(\lambda,g)\\
a^2 &= \lambda + \lambda^2P^\prime(\lambda,g) \nonumber
\end{align}
with $g$ a free parameter that tends to $\alpha$ in the continuum limit $\lambda \to 0$, and where $P$ and $P^\prime$ are multivariate polynomials. The existence of the continuum limit then requires that e.g.\ the lattice correlation length $\xi(\mu,\lambda,D)$ corresponds to a physical correlation length $\xi_p = \xi a$ that is only a function of $\alpha$ and the gap in the physical spectrum of inverse correlation lengths $\delta_p = \delta / a$, giving rise to the continuum scaling hypothesis
\begin{equation}
	\xi(\mu,\lambda,D) = \frac{1}{a} \xi_p \left( \alpha, \frac{\delta}{a} \right)
\end{equation}
with $\mu$, $\lambda$ and $\delta$ parameters in lattice units.

As the continuum theory exhibits itself a phase transition at $\alpha = \alpha_c$, the scaling hypothesis for the field theory requires that $\xi_p$, now parameterized in terms of $\Delta \alpha = \alpha - \alpha_c$ and $\delta_p = \delta/a$ is a generalised homogeneous function
\begin{equation}
	s \xi_p(\Delta \alpha, \delta_p ) = \xi_p( s^{-1/\nu} \Delta \alpha, s^{-1} \delta_p)
\end{equation}
and thus $\delta_p \xi_p(\Delta \alpha, \delta_p) = \tilde{\xi}(\delta_p^{-1/\nu} \Delta \alpha)$. Combining the IR scaling hypothesis for the critical field theory with the continuum scaling ansatz, yields a double collapse for the quantities of the lattice theory
\begin{equation}
\delta \xi(\mu,\lambda,D) = \tilde\xi \left( \left(\frac{\delta}{a} \right)^{-1/\nu} \Delta\alpha \right)
\end{equation}
For the double collapse equation of the order parameter $\braket{\phi}$ the steps are very similar, except that now we consider multiplicative corrections to the wave-function renormalization:
\begin{displaymath}
\tilde{\phi}(\Delta \alpha\bigg(\frac{\delta}{a}\bigg)^{-\frac{1}{\nu}})= (\frac{\delta}{a})^{-\frac{\beta}{\nu}}\braket{\phi}(\mu,\lambda,D)[1+\lambda P^{\prime\prime}(\lambda,g)]\,.
\end{displaymath}

The phase transition in $\lambda\phi^4$ field theory has been studied by lattice Monte-Carlo simulations \cite{Schiach2009, Bosetti2015, Bronzin2019}, hamiltonian truncation \cite{Rychkov2015} and tensor-network methods \cite{sugihara2004density,Kadoh2019, Milsted2013}, where the most accurate estimates \cite{Milsted2013, Bronzin2019} agree on a value for the critical point $\alpha_c\approx11.05-11.07$. We have run the vumps algorithm for generating $701$ data points with arbitrary lattice spacing $0.005<a^2<0.1$, bond dimensions ranging from 50-150, and couplings $\alpha$ around the critical point. Our scaling approach allows to transform those 701 data points $(\xi,\braket{\phi},\mu,\lambda,\delta)$ with the best guess of $P$, $P^\prime$ and $P^{\prime\prime}$ to points $(\tilde{\xi},\tilde\phi,\alpha,a,\delta)$, plot them according to the above collapse equations and compare them to a best guess of the scaling functions. The above described cost function is optimized for $P$, $P^\prime$ and $P^{\prime\prime}$ as well as the critical coupling $\alpha_c$ and of course the scaling functions. One could also choose to fit the critical exponents and even the particular choice of $\delta$ may be included as a fit parameter. We have chosen $P$, $P^\prime$ and $P^{\prime\prime}$ to be of order 1, 2 and 3 in $\lambda$ and order 3, 4 and 5 in $g$ respectively.
\begin{figure} \begin{center}
\subfigure{\includegraphics[width=0.99\columnwidth]{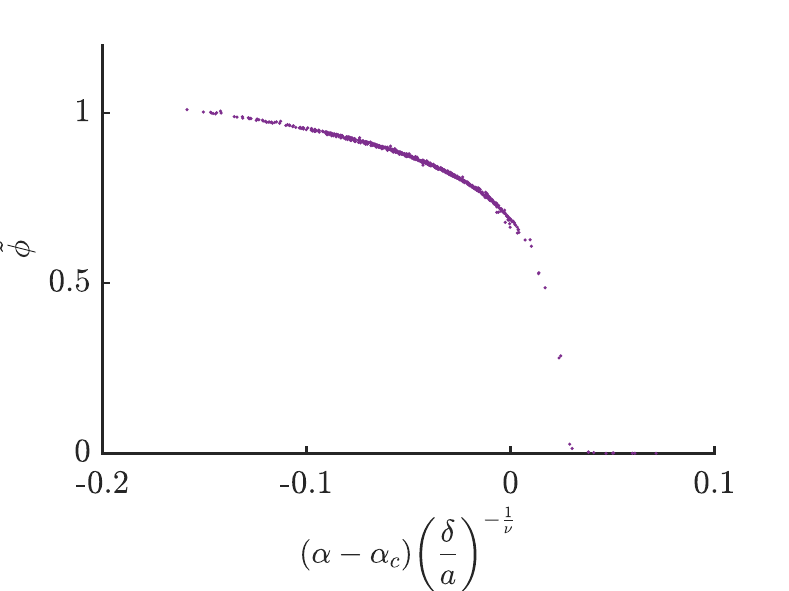}} \\
\subfigure{\includegraphics[width=0.99\columnwidth]{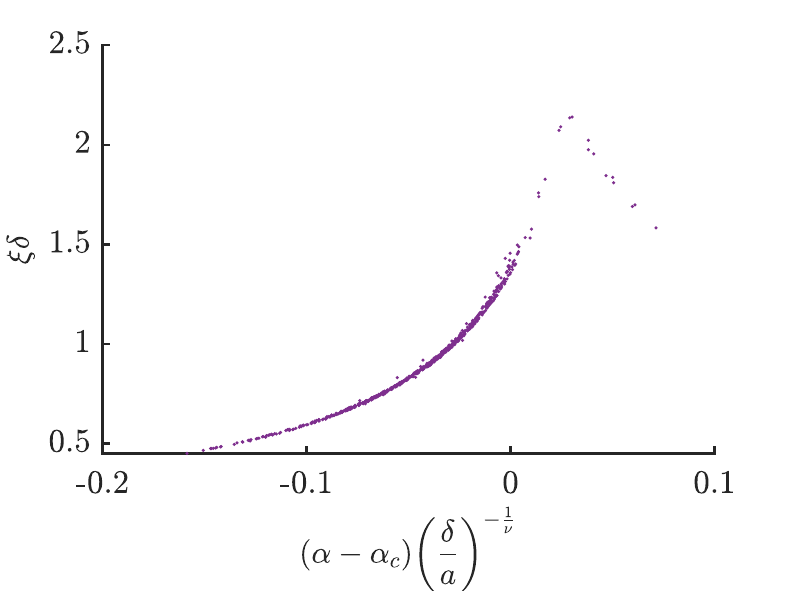}}
\caption{Double collapse plot of the order parameter and correlation length for $\lambda \phi^4$ field theory. }
\label{fig:phi4}
\end{center} \end{figure}
From the first order in $\lambda$ fit we find $\nu=1.012$, $\beta=0.130$ and $1/\alpha_c=11.0596$. If we fix $\nu=1$ and $\beta=1/8$ and fit using first, second and third order in $\lambda$ we find respectively $1/\alpha_c=11.06093$, $1/\alpha_c=11.06072$ and $1/\alpha_c=11.06886$. These should be compared to $1/\alpha_c=10.913(56)$, an alternative tensor network based study of $\lambda\phi^4$ \cite{Kadoh2019} and $1/\alpha_c=11.055(14)$, the leading MC study \cite{Bronzin2019}.

\noindent \emph{Conclusions.}  We formulated a finite scaling hypothesis for matrix product state based simulations of transfer matrices of critical classical spin systems. We identified a natural analogue of the inverse system size $1/L$ in terms of a scaling parameter $\delta\simeq 1/L$ which is a function of the eigenvalues of the transfer matrix of the MPS; for MPS simulations with finite bond dimension can be related to simulations on a halve infinite strip of size $L\times \infty$. We observed data collapses for correlation length, entanglement entropy and order parameter as a function of $\delta$ in the case of Ising, Potts and $\lambda\phi^4$ theory in $2+0$ dimensions.  Similar results would have been obtained in the quantum Hamiltonian limit, but classical spin systems were studied because of the versatility and robustness of the variational MPS algorithm for such systems. A double data collapse was obtained for data calculated for the  $\lambda\phi^4$ theory as a function of bare parameters and lattice spacing. An open question is whether a similar collapse can be obtained for the case of non-superrenormalizable field theories. Another open question arises in simulating $2+1$ and $3+0$ critical spin systems with PEPS, where similar scaling ideas lead to two inverse length scales $\delta_1$ and $\delta_2$. This situation is similar to considering a system on a $L_1\times L_2\times \infty$ cuboid for which we can borrow the scaling hypotheses used in exact diagonalization and Monte Carlo. This will be reported elsewhere.

\noindent \emph{Acknowledgments}
This work was made possible through the support of the ERC grants QUTE (647905), ERQUAF (715861) and QTFLAG.  

\bibliography{bibliography}

\end{document}